\documentclass[12pt,preprint]{aastex}

\shortauthors{Tsukamoto and Makino}

\title{Formation of protoplanets from massive planetesimals in binary systems}
\author{YUSUKE TSUKAMOTO}

\affil{Department of Earth and Planetary science,
University of Tokyo,7-3-1 Hongo, Bunkyo-ku, Tokyo 113-0033, Japan; tukamoto@margaux.astron.s.u-tokyo.ac.jp}

\and
\author{ JUNICHIRO MAKINO}

\affil{Division of Theoretical Astronomy, National
Astronomical Observatory, 2-2-1 Osawa, Mitaka, Tokyo 181-8588, Japan}

\begin{abstract}

More than half of stars reside in binary or multiple star systems and
many planets have been found in binary systems. From theoretical point of
view, however, whether or not the planetary formation proceeds in a
binary system is a very complex problem, because secular perturbation
from the companion star can easily stir up the eccentricity of the
planetesimals and cause high-velocity, destructive collisions between
planetesimals. Early stage of planetary formation process in binary
systems has been studied by restricted three-body approach with gas
drag and it is commonly accepted that accretion of planetesimals can
proceed due to orbital phasing by gas drag.  However, the gas drag
becomes less effective as the planetesimals become massive.  Therefore
it is still uncertain whether the collision velocity remains small and
planetary accretion can proceed, once the planetesimals become
massive. We performed {\it N}-body simulations of planetary formation
in binary systems starting from massive planetesimals whose size is
about 100-500 km.  We found that the eccentricity vectors of
planetesimals quickly converge to the forced eccentricity due to the
coupling of the perturbation of the companion and the mutual
interaction of planetesimals if the initial disk model is sufficiently
wide in radial distribution. This convergence decreases the collision
velocity and as a result accretion can proceed much in the same way as
in isolated systems. The basic processes of the planetary formation,
such as runaway growth and oligarchic growth and final configuration
of the protoplanets are essentially the same in binary systems and
single star systems, at least in the late stage where the effect
of gas drag is small. 

\end{abstract}


\begin{document}
\maketitle

\section{Introduction}
As of March 2007, 215 candidates of extra-solar planets have been
found and at least 30 of them are in binary or multiple star systems
(Raghavan {\it et al.} 2006). Table \ref{candidates} shows the
candidates of the close binary or multiple star system which has the
planets.  In this table, $\gamma$ Cephei is an example of
close binary systems in which we are interested.  According to Hatzes
{\it et al.} (2003), the semi-major axis and eccentricity of the
companion star of $\gamma$ Cephei are $18.5$ AU, and 0.36,
respectively. Planetary formation process in such a close binary
system is the main target of this paper.  The frequency of planets in
binary systems is not significantly different from that for single star
systems (Desidera and Barbieri 2007). It is very important to
investigate the formation process in binary system because more than
half of stars reside in binary or multiple star systems.

Many authors have investigated planetary accretion process using {\it
N}-body simulation (e.g., Aarseth, Lin, and Palmer 1993; Kokubo and
Ida 1996, 1998, 2000, 2002). In all of these simulations, the
evolution of protoplanetary disks around isolated stars was studied.
On the other hand, Quintana {\it et al.} (2002, 2006) have
investigated the planet formation in binary system from protoplanets
by {\it N}-body simulations. They studied the late stage of the
planet formation process (from protoplanets to planets).

Marzari and Scholl (2000) and Th\'ebault {\it et al.} (2004, 2006)
investigated the distribution of the collision velocity between
planetesimals with and without gas drag in binary systems by
restricted three-body approach.  They found that orbital phasing
occurred and collision velocity remained small due to the coupling of
gas drag effect and secular perturbation, and they concluded that
planetary accretion could proceed.  As the planetesimals become
massive, however, the gas drag becomes ineffective and mutual
gravitational effect between planetesimals becomes important.  In such
a condition, whether the collision velocity remains small is not
clear. As stated above, the final stage, from protoplanets to planets,
has been studied by $N$-body simulations, but there has been no $N$-body
work on the intermediate stage from massive planetesimals whose size
is 100 km - 500 km to protoplanets. In this paper, we focus on this
intermediate stage.

In a binary system, the orbits of planetesimals change due to
perturbation from the companion. The most important term of the
perturbation in our simulation region is secular perturbation, if the
orbit of the companion is eccentric. The effect of secular perturbation is
that eccentricity vector moves on the "perturbation circle". Thus, if
interaction between planetesimals and gas drag are neglected,
planetesimals with different semi-major axis move on the perturbation
circles on different frequencies and phases, and therefore gain high
relative velocity. This is the reason why the collision velocity becomes high
in previous works without gas drag or self gravity (Marzari and Scholl 2000;
Th\'ebault {\it et al.} 2004, 2006). If the collision velocity becomes
high when the planetesimals become massive and thus, the effect of gas drag
becomes small, accretion process might halt since collisions might be
destructive.  

It is a very important question whether destructive collisions occur when
the gravitational interaction between planetesimals is taken into account 
because it determines whether the accretion process can continue after
gas drag becomes ineffective. Ito and Tanikawa (2001) studied the
evolution of the  orbital elements of protoplanets under the
perturbation of Jupiter and found that the eccentricities of the
protoplanets aligned with each other, resulting in the evolution was
very similar to that without the influence of Jupiter.  This alignment
is due to gravitational interaction between the protoplanets. If this
kind of alignment also occur for planetesimals in binary system,
the collision velocity can become smaller than the value predicted by
Marzari and Scholl (2000). 

In this paper, we report the results of {\it N}-body simulations of
protoplanet formation from massive ($3 \times 10^{23} - 1.4 \times
10^{24}$ g) planetesimals. We start simulations with the
planetesimal disk in which all planetesimals have the same mass. The
gravitational interaction between planetesimals is included and 
gas drag is neglected. We summarize theory of secular perturbation
and discuss what initial distribution should be used in section 2.
The numerical scheme and initial conditions are described in section
3.  We also discuss the validity of the "same mass" assumption of
planetesimals in section 3. The results of {\it N}-body simulations
are presented in section 4. In section 5, we sum up. 

\begin{table}[hbt]
\
\caption{Known candidates of close systems}		
\label{candidates}
\begin{center}
\begin{tabular}{cccccccc}
\hline\hline
  Name 
& Component
& \begin{tabular}{c} $ a \sin i $ \\(AU) \end{tabular} 
& \begin{tabular}{c} Projected \\ Separation \\ (AU) \end{tabular} 
& \begin{tabular}{c} $M \sin i$ \\ ($M_{J}$)\end{tabular} 
& eccentricity 
& References \\
\hline

 HD 041004 & B & ... & 22  & ...  & ... & 1, 2, 3, 4  \\
           & C & 0.016 & ... & 18.4 & 0.08 & 3, 4  \\
           & b & 1.31 & ... & 2.3 & 0.39 &  \\
\hline

 $\gamma$ Cephei & B & 20.3 & ...  & ...  & 0.39 & 1, 5, 6, 7, 8  \\
                 & b & 2.03 & ... & 1.59 & 0.2 &  \\
\hline

 GJ 893  & B & ... & 2248  & ...  & ... & 9, 10  \\
        & C & ... & 18    & ... & ... &   \\
        & b & 0.3 & ... & 2.9 & ... &  \\
\hline\hline

\end{tabular}
\end{center}

\tablecomments{These parameters are from Raghavan {\it et al.} (2006) and
        its references. Column component lists companions (B, C,
        D,...) and planets (b, c, d, ...).}
\tablerefs{(1)Eggenberger {\it et al.} 2004;
(2)See 1896; (3) Zucker {\it et al.} 2003; (4) Zucker {\it et al.}
        2004; (5) Mason {\it et al.} 2001; (6) Campbell {\it et al.}
        1988; (7) Griffin {\it et al.} 2002; (8) Hatzes {\it et al.}
        2003; (9) Zacharias {\it et al.} 2004; (10) Wilson 1953}
\end{table}

\section{Theoretical Preparation }
\subsection{Brief description of secular perturbation}
If the mutual interaction of planetesimals and gas drag are
negligible, the orbital evolution of a planetesimal in a binary system
can be described by restricted three-body approximation, and the time
evolution of eccentricity and longitude of pericenter of the planetesimal
are given by secular perturbation theory.  The secular evolution of
the {\it h} and {\it k} variables of the planetesimal are expressed as
(Heppenheimer 1978; Whitmire {\it et al.} 1998; Th\'ebault {\it et
al.} 2006) 
\begin{eqnarray}
h(t)& = &e_{p} \sin (At+\varpi_{0})~,\\
k(t)&= &e_{p} \cos (At+\varpi_{0}) + e_{f} ~,
\end{eqnarray}
where
\begin{eqnarray}
h(t)& = &e \sin ( \varpi)~,\\
k(t)&= &e \cos ( \varpi) ~, 
\end{eqnarray}
and $e$  and $ \varpi$ are the eccentricity and the longitude of
pericenter of the planetesimal. Here, {\it A}, $e_{p}$, $ \varpi_{0}$, and
$e_{f}$ are constants which depend on the strength of the secular
perturbation. We take the {\it k} axis as direction of the
eccentricity vector of the companion. In other words, the eccentricity
vector of the companion is $( e_{B}, 0 )$. The semi-major axis of the
companion is $a_{B}$ and its mass is $M_{B}$. According to the
secular perturbation theory, $e_{f}$ is given by  
\begin{eqnarray}
e_{f} = \frac{5}{4} \frac{a}{a_{B}} \frac{e_{B}}{(1-e^{2}_{B})} ~.
\end{eqnarray}
This $e_{f}$ is forced eccentricity induced by the companion
star. Here, 
{\it a} is the semi-major axis of the planetesimal.  The angular
velocity of rotation on k-h plane, {\it A}, is given by  
\begin{eqnarray}
A = \frac{3}{2} \pi \frac{1}{(1-e^{2}_{B})^{3/2}}m_{B} \frac{a^{3/2}}{a_{B}^{3}} ~.
\end{eqnarray}
We use system of units in which the solar mass, 1 AU and 1 year
are all unity.  Remaining two constants, $\varpi_{0}$ and $e_{p}$, are
determined from the initial orbital elements of the planetesimal. We
call the vectors $ {\bf  e } = (k,h),~ {\bf e}_{f} = (e_{f} , 0) $ the
eccentricity vector and the forced eccentricity vector,
respectively. On k-h plane, the eccentricity vector of a planetesimal
moves on a circle centered at ${\bf e}_{f}$.

\subsection{Initial eccentricity of planetesimals}
We consider two types of initial conditions for planetesimals. In the
first one, the initial distribution of  eccentricity vectors of
planetesimals is centered at the forced eccentricity vector. The
distributions of ${\bf e} -{\bf e}_{f}$ and inclination {\it i} are
both given by Rayleigh distribution with dispersions  $\langle  |{\bf
e}- {\bf e}_{f}|^{2} \rangle ^{1/2}=2 \langle i^{2} \rangle
^{1/2}=0.02$. We call this model "forced" model. In the second model,
the distribution of eccentricity vector is centered at the origin of 
k-h plane. The distributions of eccentricity {\it e} and inclination
{\it i} are also given by Rayleigh distribution with dispersions
$\langle e^{2} \rangle ^{1/2}=2 \langle i^{2} \rangle
^{1/2}=0.02$. This is the same distribution as those used in previous works
for {\it N}-body simulation of planetary formation in isolated systems
(Kokubo and Ida 2002).  We call this model "circular" model. 

We argue that the "forced" model is more suitable for simulation
in binary system than the circular model for the following
reasons. Consider a narrow region of a planetesimal disk such as the
region of 0.95 AU $< a <$ 1.05 AU.  If the eccentricity vectors of
planetesimals in this region do not distribute around the forced
eccentricity vector, they will rotate around the position of the
forced eccentricity vector.  The planetesimals in a narrow region
would rotate together because the angular velocity {\it A} is similar
for planetesimals with similar values of the semi-major axis {\it
a}. In addition, they align due to secular interactions between
planetesimals (see, e.g., Ito and Tanikawa 2001),  Because of this
collective motion, the relative velocity between planetesimals is kept
small in this narrow "ring".   
 
However, if we consider a wider region, it would behave as collection
of many narrow rings. If we ignore interactions between the rings, they
would rotate on their own angular velocities {\it A}. Since
neighboring rings have slightly different values of {\it A}, they
would soon physically collide with each other, resulting in
damping of "free" eccentricity.  Gravitational interaction between
rings would also dump relative difference of eccentricity vectors.
In other words, planetesimals in the
circular model would first relax to the forced model. We argue that
the equilibrium state is more suitable for the initial conditions.
This is why we consider "forced" model. 

As we mentioned in introduction, we study the late stage of the
formation process of protoplanets. The earlier phase have been studied  
by Marzari and Scholl (2000) and others. Thus, it might seem reasonable 
to set the initial
eccentricity and longitude of pericenter of planetesimals  to the
equilibrium value of the orbital phasing in Marzari and Scholl 
(2000) instead of the circular model. 
However, We found it is a bit  difficult to use
their results, because they adopt circular gas disk model. This circular gas
disk in a binary system with eccentric companion seems a bit
unnatural. Due to the perturbation 
from the companion star, the gas disk may be twisted. We could not
find previous works which directly studied the equilibrium state of
gas disk under the secular perturbation of companion, but recent work
by Papaloizou (2005) seems to imply that eccentric disk can be
long-lived even if there is no companion. So it seems likely that gas
disk is not circular when eccentric companion exists. This may change
gas drag effect to the planetesimals and might affects the
direction of the orbital phasing. This is the reason why we do not
use the Marzari's results. Even if the orbital elements of them is
correct, our circular model is very close to their results and it would
soon relax to the "forced" model.  So we expect that it  makes no significant
change to our results.  Clearly, more studies on the dynamics of gas
disk is necessary.

\subsection{Collision velocity}
As noted in section 2.1, we adopt the Rayleigh distribution with
dispersions $\langle e^{2} \rangle ^{1/2}=2 \langle i^{2} \rangle
^{1/2}=0.02$ for the "circular" model (Ida and Makino 1992), and $\langle
|{\bf e}- {\bf e}_{f}|^{2} \rangle ^{1/2}=2 \langle i^{2} \rangle
^{1/2}=0.02$ for the "forced" model (see Table 2 ). These conditions imply
collision velocity is initially  $v_{col} \simeq 500-1000~
\mathrm{m~s^{-1}}$ at 1 AU for the initial distribution. When the
effect of gas drag is taken into account, the equilibrium  eccentricity
and inclination are estimated as $ \langle e^{2} \rangle ^{1/2}=2
\langle i^{2} \rangle ^{1/2}=0.0042$ (Kokubo and Ida 2000).  This
value is probably more reasonable.
In this paper, however, we use the value 0.02
because we neglect gas drag force in our calculation and the value of
the eccentricity and inclination will soon relax to the value of gas-free
case. Our focus is not on the quantitative analysis of the collision
velocity but on the qualitative analysis of the coupling of
secular perturbation and gravitational interactions between
planetesimals. To determine the realistic value of the collision
velocity, we should include gas drag force in our calculation. 

\section{ Method of Calculation}
\subsection{Initial conditions}
The model parameters for the companion star are summarized in table
\ref{p of c}. Mass of the primary star is $ 1~ M_{ \odot } $ for all
models. We adopt three models for companion. Model alpha corresponds to
$\alpha$ Cen system. The mass ratio is chosen to be the same as that
of $\alpha$ Cen system, (1.1:0.9). For the other two models, we use a
somewhat smaller semi-major axis than those of observed binary systems in
which the planets are found. We choose this value to study the case
in which secular perturbation is strong.  For circular models, we use
an axisymmetric surface mass density distribution for planetesimal
disk whose surface mass density is given by  
\begin{eqnarray}
\Sigma_{solid} = \Sigma_{0}(\frac{a}{ \mathrm {1~AU} } )^{-3/2}
~\mathrm{ g~ cm^{-2} } ~,
\end{eqnarray}
where {\it a} is distance from the primary star, and $ \Sigma_{0}$
is the reference surface density at 1 AU (We adopt $ \Sigma_{0} = 10
~\mathrm{g ~cm^{-2}}$). Table \ref { inicond } shows model
parameters. Here, {\it N} is the number of planetesimals.  For
radial distribution, we use two disk models, "wide disk" models (models
0 - 5) and "narrow disk" models (models 6 - 8).   In the "wide disk"
models, we set inner and outer cutoff radii to be 0.5 AU and 1.5 AU,
respectively. In the "narrow disk" models, inner and outer cutoff radii
are 0.95 AU and 1.05 AU, respectively. Planetesimals have equal mass
in all models. The number of planetesimals is 10000  in the wide disk
models. In two narrow disk models (model 6, 7), Number of
planetesimals is 5000. In one of narrow disk models (model 8), the
number of planetesimals is 975.  This is a "cutoff" model which has the
planetesimals of the same mass as in the wide disk models. The density of
planetesimals is 2 $\mathrm{g~cm^{-3}}$. We increase their radii by a
factor 5 to accelerate accretion process (Kokubo and Ida 1996). At
first, all planetesimals have same mass ($ \simeq 1.44 \times 10^{24}$
g in models 0 - 5 and 8 and $2.88 \times 10^{23}$ g in model 6 and 7).
 
One critical question is if our "same mass" setup can be really
regarded as description of the later stage of planetary
formation. Recent theoretical and numerical works of planetary
accretion (for example Inaba {\it et al.} 2001, Rafikov 2003) seem to
suggest that the "orderly" phase of growth does not exist, and runaway
growth starts from much smaller mass than that of our setup.  Our
setup, therefore, is not realistic. However, statistical calculation by
Inaba {\it et al.} (2001)  has shown that once massive
planetesimals ($ M \simeq 10^{23}$ g) formed, planetesimals with mass
less than $10^{21}$ g become dynamically unimportant. So even though
our initial condition is oversimplified, it might still give
qualitatively valid description of the later phase. 

Two planetesimals are considered to collide when their distance
becomes less than the sum of their radii. We assume the perfect
accretion under which planetesimals always accrete when they
collide. Whether or not this assumption is good depends on the
distribution of the collision velocities.  The escape velocity of our
planetesimals is $200-600 ~\mathrm{m~s^{-1}}$ and is of the same order
as the initial collision velocity. Furthermore, if gas effect is
included, the eccentricity 
dispersion might be less than our estimate and the collision velocity
becomes smaller. So we believe the assumption of perfect accretion is
valid.

As stated above, we increase the radii of planetesimals {\it f}-fold
(here, $f=5$) to save calculation time. This acceleration of
accretion may affect time evolution of planetesimals, especially
the time evolution of orbital elements on k-h plane because the
increase of radii changes timescale of accretion but does not
affect secular perturbation of companion star.

\begin{table}[hbt]
\
\caption{Model parameters for the companion star}		
\label{p of c}
\begin{center}
\begin{tabular}{cccc}
\hline\hline
 Name & Mass & \begin{tabular}{c}semi-major axis \\(AU) \end{tabular} &
 eccentricity \\
\hline

 nocomp & 0 & .... & .... \\
 e25 & $ 1 M_{ \odot }$ & 16  & 0.25 \\
 e50 & $ 1 M_{ \odot }$ & 16 & 0.5 \\
 alpha & $ 0.82 M_{ \odot } $ & 23.4 & 0.52 \\
\hline

\end{tabular}
\end{center}
\end{table}

\begin{table}[hbt]

\caption{Initial conditions}		
\label{ inicond }
\begin{center}
\begin{tabular}{ccccccc}
\hline\hline
 Model & {\it N} &
 \begin{tabular}{c} Width of disk \\ (AU) \end{tabular}
&  companion & initial orbit 
\\
\hline

 0 & 10000 & 0.5-1.5 & nocomp & ...   \\
 1 & 10000 & 0.5-1.5 & e25 & forced   \\
 2 & 10000 & 0.5-1.5 & e50 & forced   \\
 3 & 10000 & 0.5-1.5 & e25 & circular   \\
 4 & 10000 & 0.5-1.5 & e50 & circular   \\
 5 & 10000 & 0.5-1.5 & alpha & circular   \\
 6 & 5000 & 0.95-1.05 & e25 & circular  \\
 7 & 5000 & 0.95-1.05 & alpha & circular  \\
 8 & 975 & 0.95-1.05 & e25 & circular   \\

\hline

\end{tabular}
\end{center}
\end{table}

\subsection{Integration scheme}
We use the fourth-order Hermite scheme (Makino and Aarseth 1992) with
hierarchical timesteps (Makino 1991) improved for planetary systems
(Kokubo {\it et al.} 1998 ) for numerical integration of planetesimals
and companion star. The equation of motion for planetesimals is
given by
\begin{eqnarray}
{\bf a}_{i} = - \sum_{i \neq j} G m_{j} \frac{ {\bf r_{i j}} }{r_{i j}^{3}} 
- G(m_{i} +M_{p})\frac { {\bf r_{i p}} } {r_{i p}^{3}} -G M_{c} 
( \frac { {\bf r_{i c}} }{ r_{i c}^{3} } + \frac { {\bf r_{p c}}
}{r_{p c}^{3}} ) ~,
\end{eqnarray}
where $M_{p},M_{c}, r_{ip}$, and $r_{i c}$ are mass of the primary
star, mass of companion star, position of planetesimal  relative
to the primary star and that relative to the companion star. We use
position of the primary star as origin of the coordinate for
planetesimals. The motion of the primary star due to the gravitational
forces of planetesimals is neglected. 

Most expensive part of the numerical integration is calculation of
mutual gravitational interaction between planetesimals. We use
GRAPE-6 (Makino {\it et al.} 2003)  and GRAPE-6A (Fukushige {\it et
al.} 2005) to calculate the gravitational interaction between
planetesimals.  We also integrate orbit of the companion star
using the fourth-order Hermite scheme. For both the planetesimals and
the companion stars, we use the standard timestep criterion (Aarseth 1985) 
\begin{eqnarray}
\Delta t =\sqrt{ \eta \frac{|a||a^{(2)}|+ |\dot{a}|^{2}}
       {|\dot{a}||a^{(3)}|+ |a^{(2)}|^{2}}} ~.
\end{eqnarray}
Since the orbital period of the companion is long, the accuracy of its
orbit is more than enough. 

\section{Results}
\subsection{Planetary accretion in binary systems}
\begin{figure}
\plotone{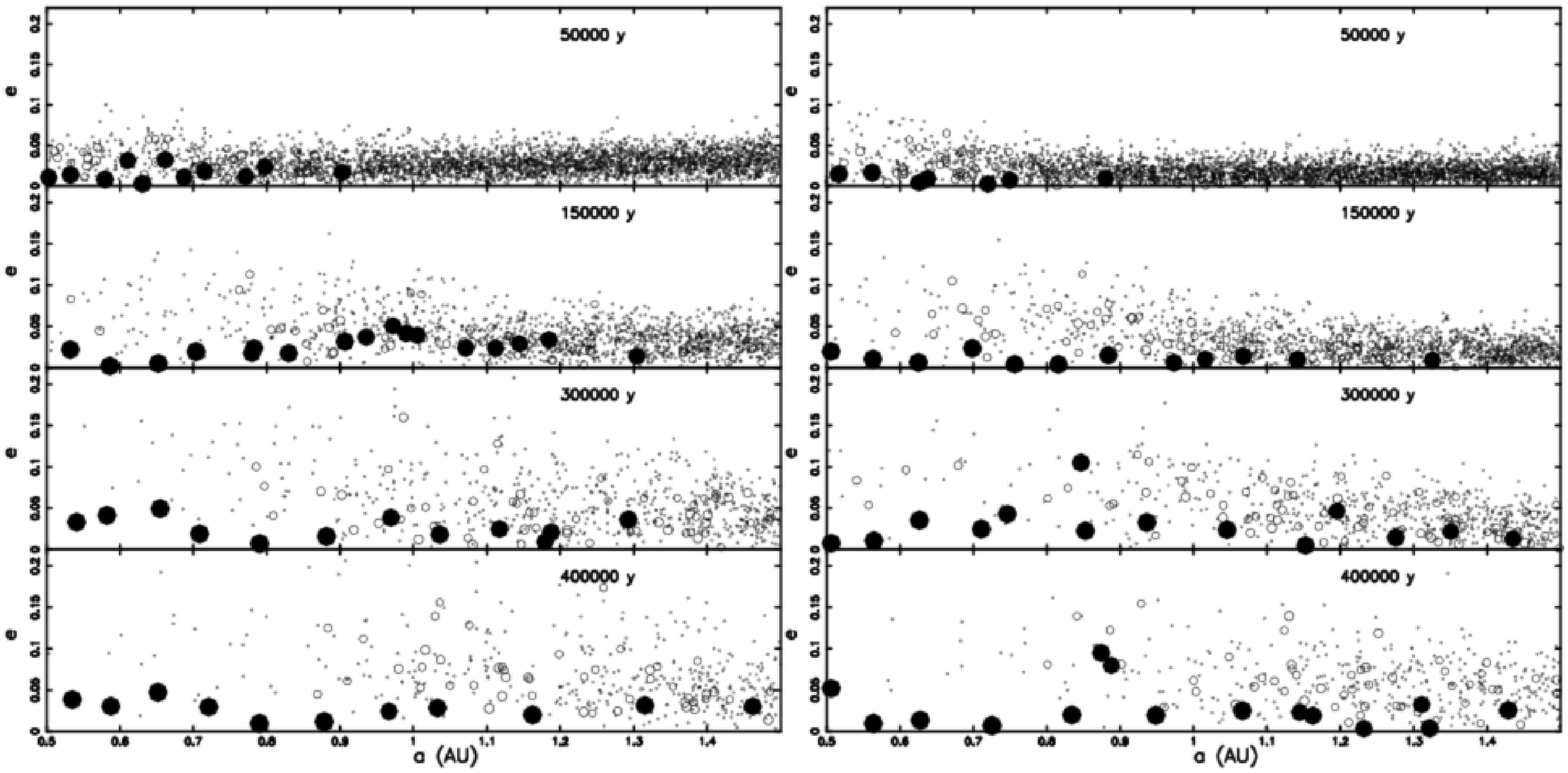}

\caption{Distribution of planetesimals on the {\it a-e} plane for model 1
(left)
and model 0 (right). The circles represent planetesimals. The filled circle represent
protoplanets with mass more than 100 times the initial mass.}
\label{snapshots}
\end{figure}

Figure \ref {snapshots}  shows the time evolution of planetesimals of
model 1 (left) and model 0 (right) on the a-e plane.  We integrate the
system for $5 \times 10^{5}$ years. We can see that  protoplanets grow
in a very similar way in these two models. This behavior is essentially
the same for all models.

\begin{figure}
\plotone{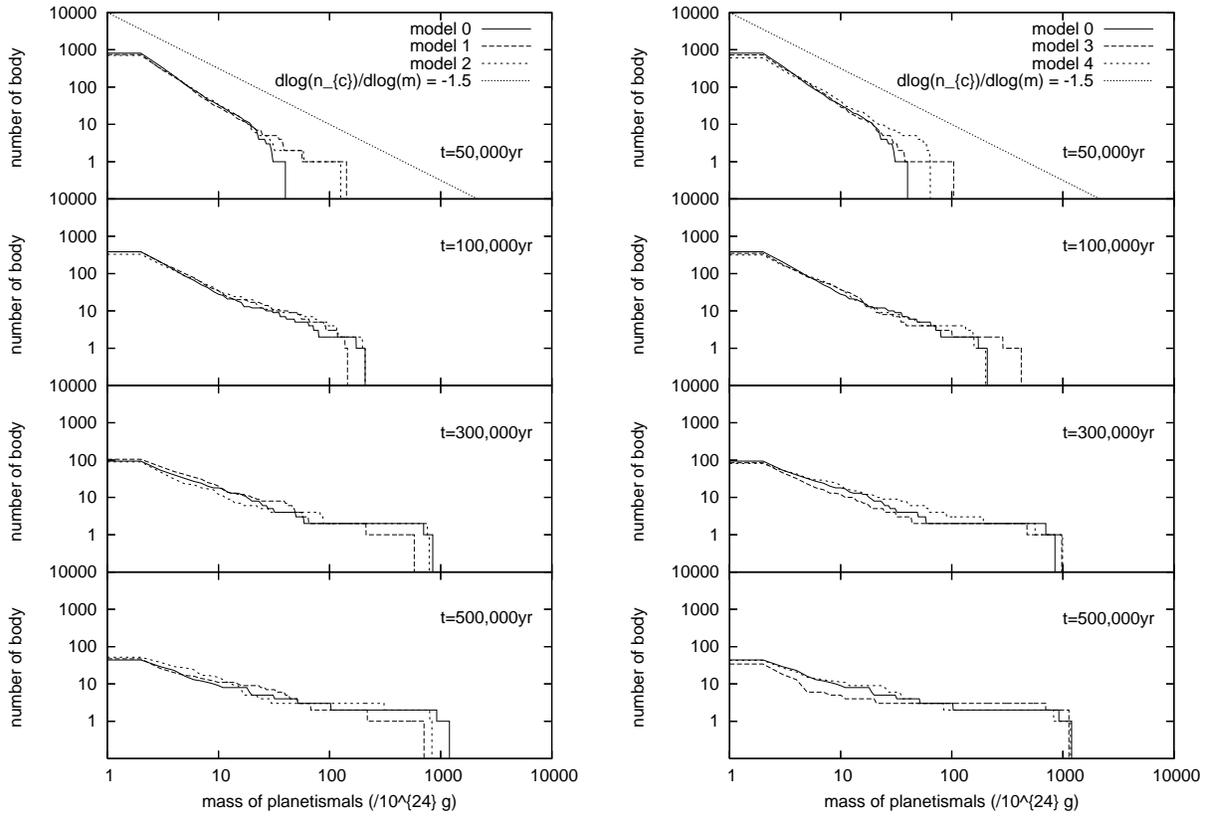}

\caption{ Cumulative mass distribution of planetesimals. Left panels
show the results of the forced models (model 1, 2) and  right panels
show the results of the circular models (model 3, 4). In all panels,
the results of isolated model (model 0) are also shown. Times are  $
5\times 10^{4},~1\times 10^{5},~ 3\times 10^{5},~5 \times 10^{5}$
years from top to bottom. The dotted lines at the top panels indicate
the slope of -1.5.} 
\label{mass_number}
\end{figure}

Figure \ref {mass_number} shows cumulative mass distribution of
planetesimals of the region $0.9 {\rm AU}< a <1.1 {\rm AU}$. The time
evolutions of all of these models are very similar. In all cases, the mass
distributions first relax to the power-law distribution with power
index $d \log n_{c}/d \log m \simeq -1.5$ (top panels) where $n_{c} $
is the cumulative number of planetesimals. This power-law distribution
is characteristic of the runaway growth (Makino {\it et al.} 1998,
Kokubo and Ida 2000).
			 
As time goes on, the mass distributions become bimodal: planetesimals
with mass $ M \simeq 1-10 \times 10^{24}$ g  and large protoplanets
with mass $ M \simeq 10^{27}$ g. This bimodal mass distribution is the
result of the runaway and the oligarchic growth of protoplanets
(Kokubo and Ida 1996,1998).
The time evolution of the mass distribution of binary systems is
almost the same as that in  single star systems. It means that
secular perturbation does not change the process of the protoplanet
formation. Runaway growth and oligarchic growth also occur in
binary systems in a similar way as in the single star system.   
 
\begin{figure}
\plotone{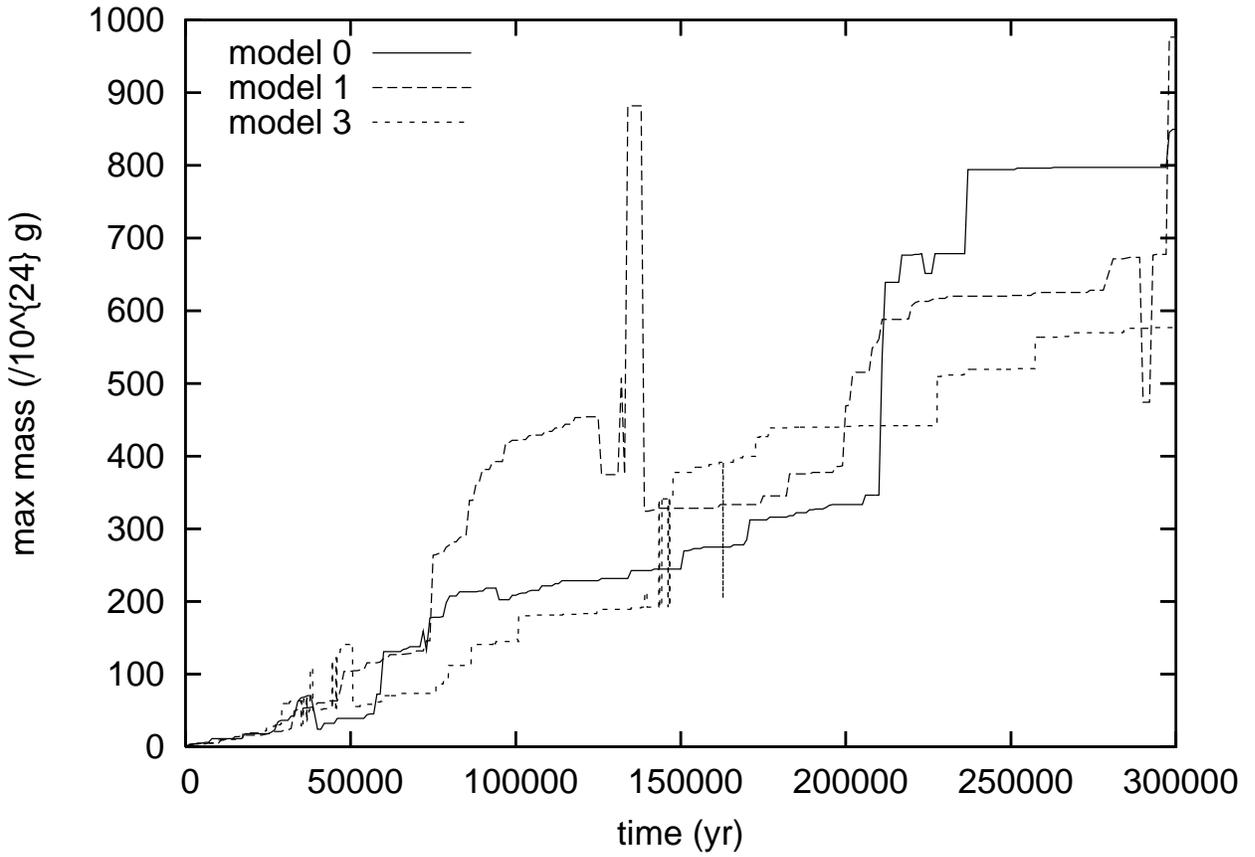}

\caption{The time evolution of maximum mass of planetesimals
between $0.9{\rm AU}<a<1.1{\rm AU}$. Solid, dashed, and  dotted
curves show the result of the model 0, 1, and 3, respectively. }
\label{time_max_e25}
\end{figure}

\begin{figure}
\plotone{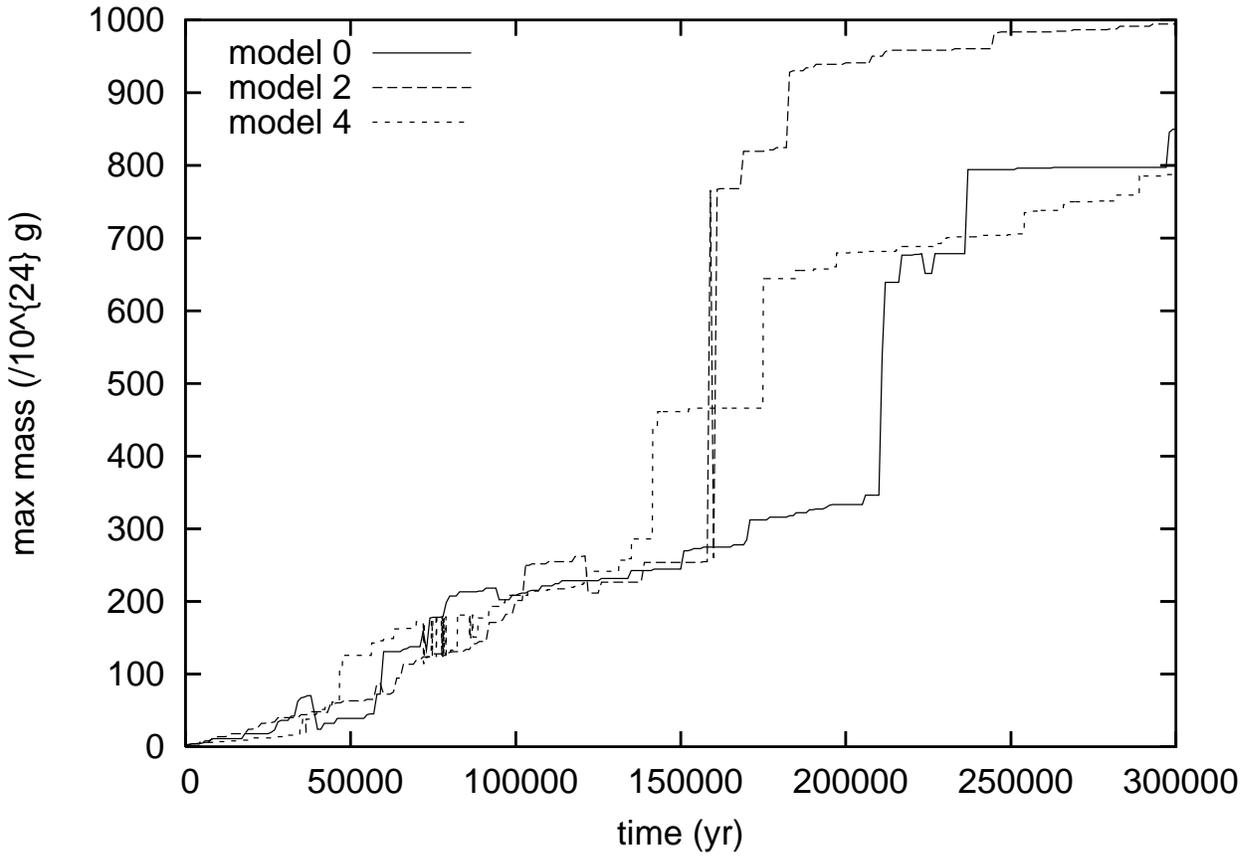}

\caption{Same as figure \ref {time_max_e25} but for models 0, 2, and 4. }
\label{time_max_e50}
\end{figure}

\begin{figure}
\plotone{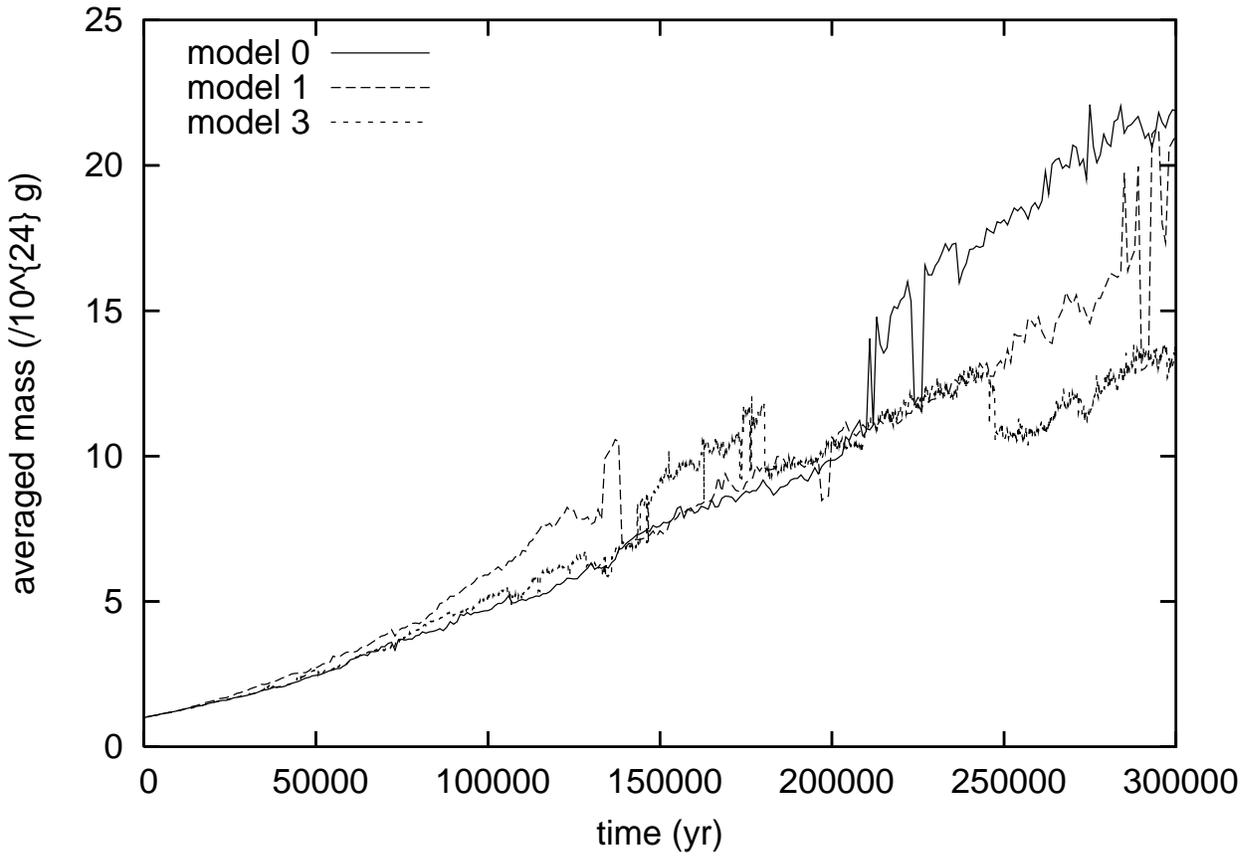}

\caption{Same as figure \ref {time_max_e25} but the average mass is shown. }
\label{time_ave_e25}
\end{figure}

\begin{figure}
\plotone{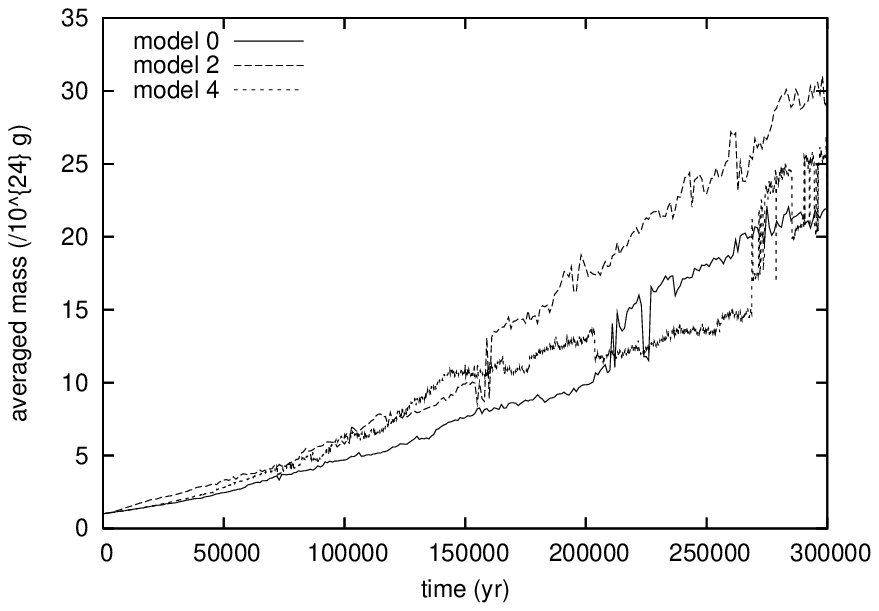}

\caption{ Same as figure \ref {time_ave_e25} but for models 0, 2, 4. }
\label{time_ave_e50}
\end{figure}

In figure \ref {time_max_e25} - \ref {time_ave_e50}, we show the
evolution of the mass of most massive planetesimals (\ref
{time_max_e25} and \ref {time_max_e50}) and average mass of
planetesimals (\ref {time_ave_e25} and \ref {time_ave_e50}).  We can
see that the evolution is rather similar and there is no systematic
tendency due to the presence of the companion.
			
\begin{figure}
\epsscale{0.8}
\plotone{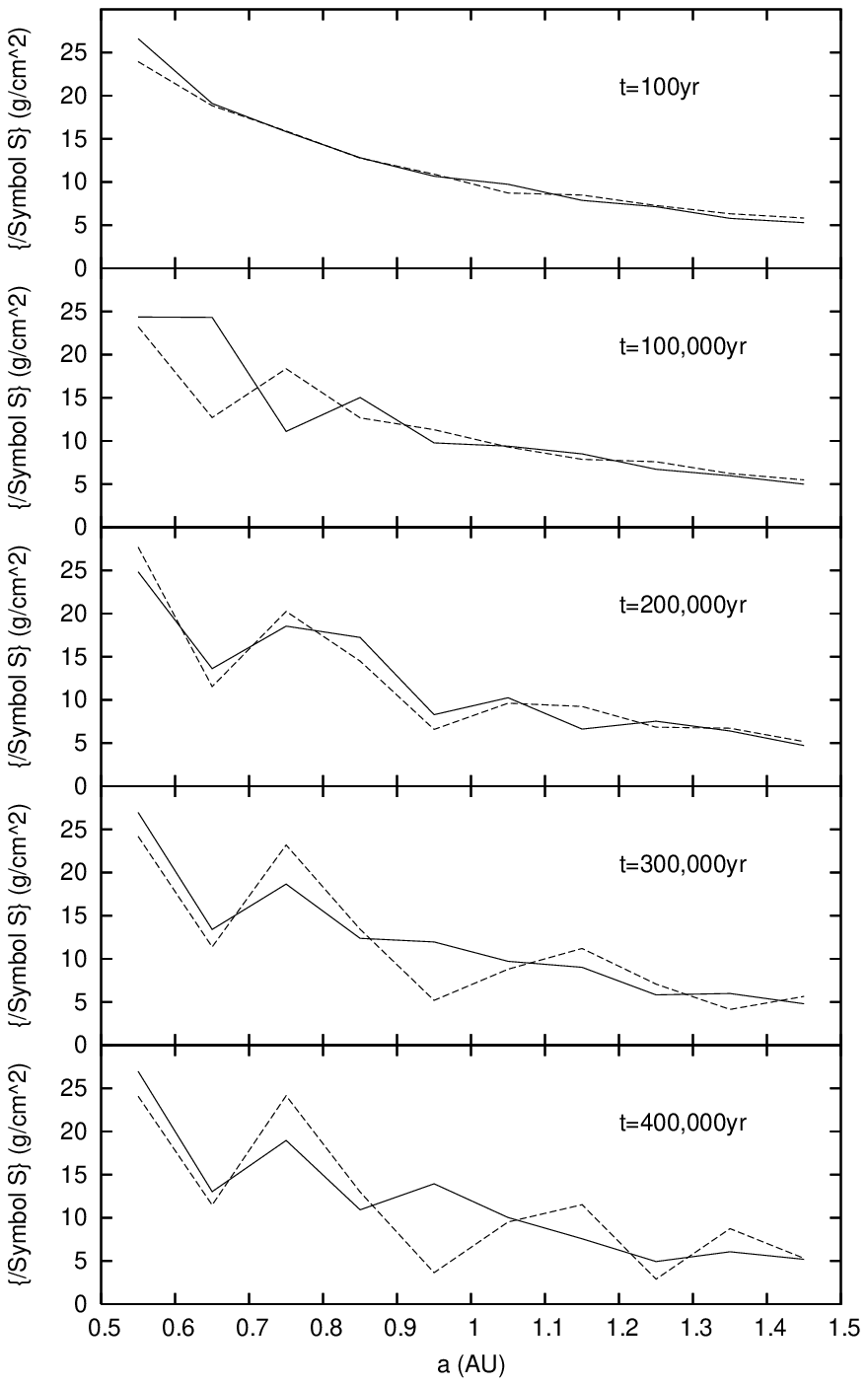}
\caption{The surface mass density of planetesimals for model 1
(solid) and model 0 (dotted) at $t=1 \times 10^{2},1
\times 10^{5},2 \times 10^{5},4 \times 10^{5}$ years, from top to bottom. }
\label{surface}
\end{figure}

The evolution of the surface mass density of model 1 ($e=0.25$,
forced) and that of model 0 (no companion)  are shown in Figure
\ref{surface}. We found that density profile of planetesimals in a
binary system is virtually the same as that in a single star system. 

From secular perturbation theory, time evolution of
semi-major axis of a planetesimal is given by  
\begin{eqnarray}
\frac{da}{dt} = \frac{2}{na} \frac{ \partial R}{ \partial \epsilon}~,
\end{eqnarray}
where $ \epsilon$ and $R$ are mean longitude at epoch and 
disturbing function, respectively. In practice, variation of
$\epsilon$ can usually be neglected since it is a small effect. Thus,
mass migration does not occur since secular perturbation hardly
change the semi-major axis of planetesimals. Our results are
consistent with this theoretical expectation.		 

\subsection{ Time evolution of eccentricity}
\begin{figure}
\epsscale{0.6}
\plotone{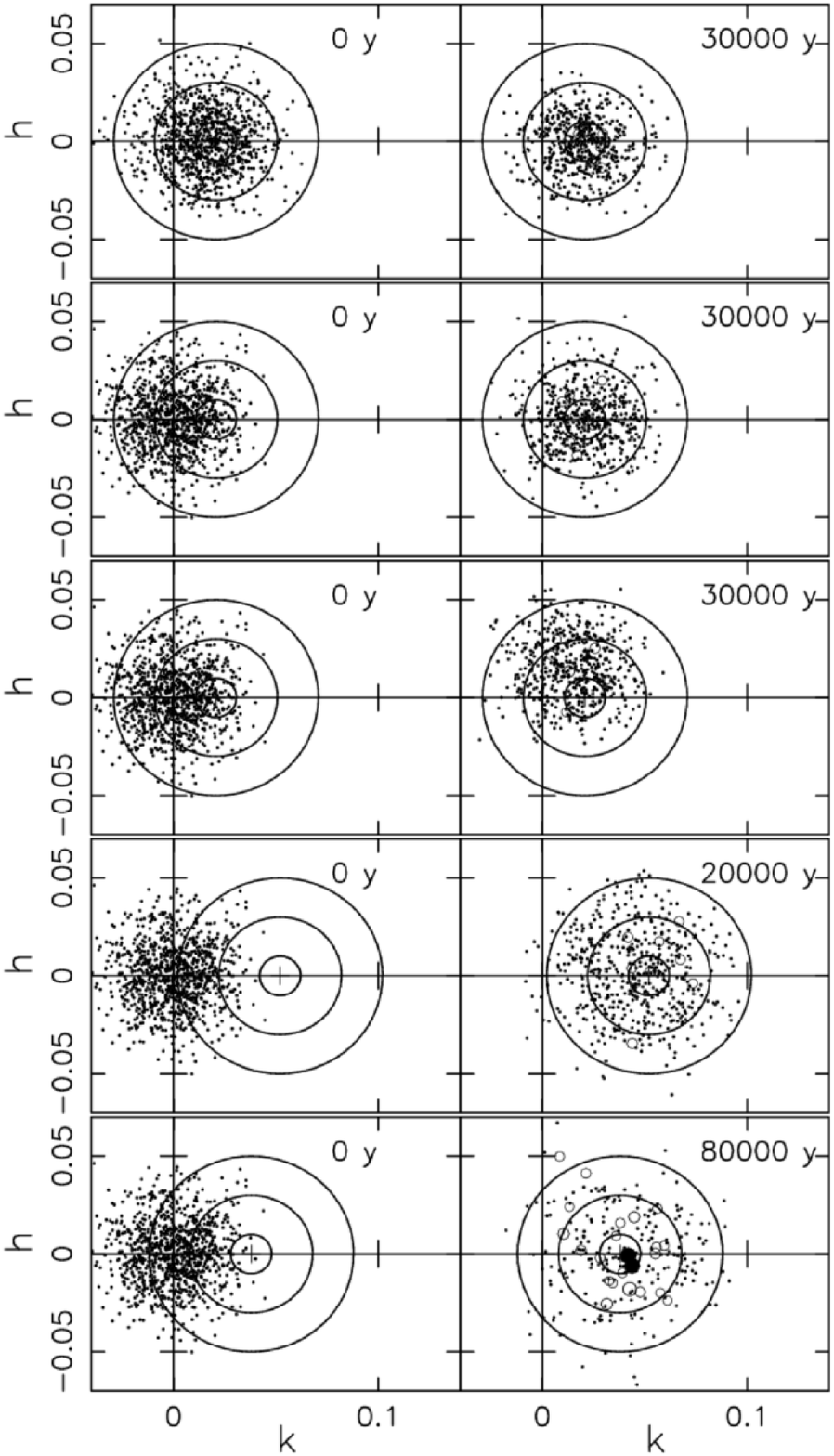}
\caption{The distribution of planetesimals of model 1, 3, 8, 4, 5 (top
to bottom) on k-h plane. Left panels show initial distributions and
right panels show evolved state ($3 \times 10^{4}$ years, for models 1,
3, and 8 and $2 \times 10^{4}$ and $8 \times 10^{4}$ years for models 4
and 5) 
We plotted planetesimals  between $0.95 ~\mathrm{AU} < a < 1.05~\mathrm{AU}$. The center of
circles is the forced eccentricity for $a= 1$ AU and radii of circles are 0.01,
0.03, 0.05 respectively.
The open and filled circles represent the planetesimals with mass 10
times and 100 times the initial value.}
\label{snapshot_hk}
\end{figure}

The time evolutions of the distribution of the planetesimals on k-h
plane of model 1 (forced model of $e=0.25$), 3 (circular model of
$e=0.25$), 8 (cutoff model of $e=0.25$), 4, and 5 (circular models)
are shown in Figure \ref {snapshot_hk}.  

The top three panels show models 1, 3, 8, all with an e25 companion. In
model 1 (top panel),  planetesimals are initially distributed around
the forced eccentricity, and this does not change in time. In model 3,
planetesimals are initially distributed around zero eccentricity, but
this initial distribution is replaced by a distribution centered at
the forced value. The behavior of planetesimals in model 8 is very
different. The distribution after 30,000 years is not centered at the
forced value and keeps rotating around the forced eccentricity vector
(see figure \ref {time_e_e25_alpha_multi}  and description below).
This difference between narrow disk models and wide disk models will be
discussed in more details in section 4.2.1.  The bottom two panels of
figure \ref {snapshot_hk} show that the eccentricity vectors of the
planetesimals of other circular models with wide distribution also
converge to the forced eccentricity vector. We will discuss the time
evolution of proper eccentricity in section 4.2.2.

\subsubsection{Difference between wide and narrow models}
We performed simulations of the narrow disk models  (models 6, 7, 8) to
see the difference between the wide and narrow distributions.  We use two
types of narrow disk models. One has the planetesimals of the same
mass as in wide disk models, and thus the number of planetesimals is small
($N=975$, model 8).  The other has a large number of smaller
planetesimals (see Table 2).   
		     
\begin{figure}
\epsscale{1.0}
\plotone{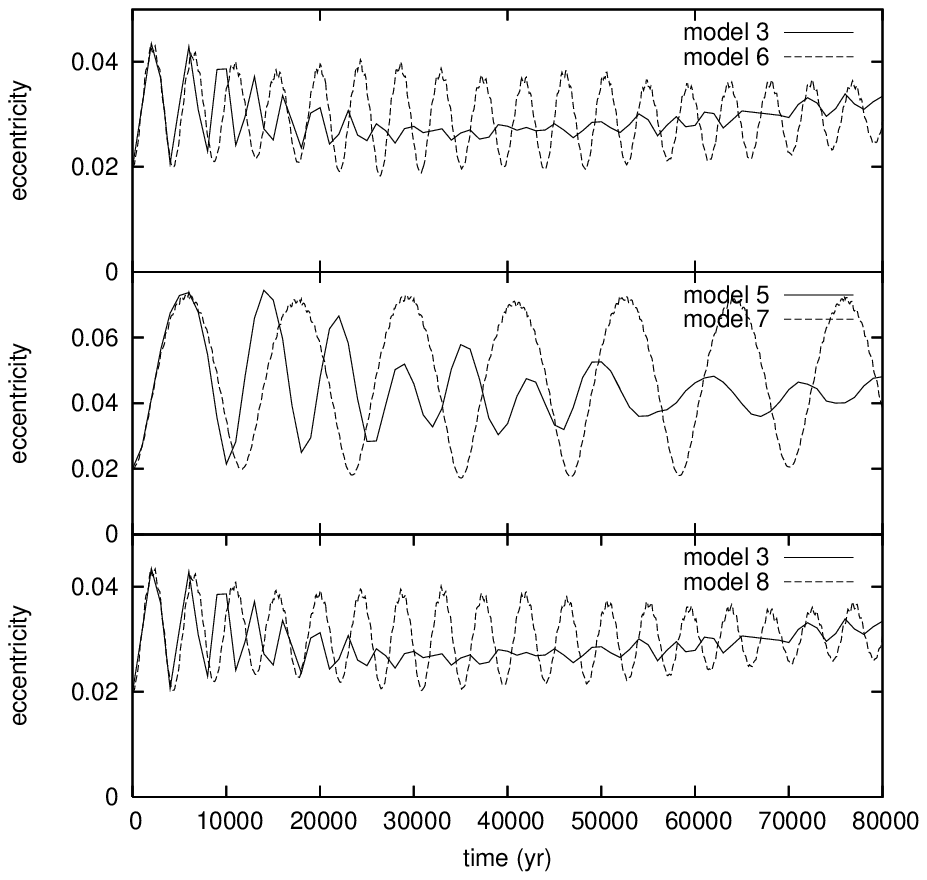}

\caption{The time evolution of the average eccentricity in  narrow disk
models (models 6 - 8) compared with those in wide disk models (models
3, 5, and 3 from top to bottom). For wide disk models, the
eccentricity is averaged for planetesimals in the range of $0.95
~\mathrm{AU} < a < 1.05 ~\mathrm{AU}$.  The solid curves show results
of wide disk models and  dashed curves show results of narrow disk models.}

\label{time_e_e25_alpha_multi}

\end{figure}

Figure \ref {time_e_e25_alpha_multi} shows the time evolution of
eccentricity.  In wide disk models,  the oscillation of eccentricity
is damped quickly due to the convergence to the forced eccentricity
vector as shown in figure \ref {snapshot_hk}. In narrow disk models,
however, the eccentricity liberates around the forced eccentricity and
does not converge to the forced eccentricity.

This difference can be understood as follows. As we discussed in
section 2, the planetesimals rotate around the forced eccentricity
vector on k-h plane and the angular velocity {\it A} is a function of
semi-major axis of planetesimals. In the case of narrow disk models,
the range of {\it A} is small, and it is possible that all
planetesimals synchronize due to mutual  gravitational
interaction. Thus in narrow disk models planetesimals move
collectively on k-h plane. In the case of the wide disk models,
however, such collective motion is not allowed since the range of {\it
A} is too large. If planetesimals with different values of {\it a}
(therefore, {\it A}) rotate around the its own values of forced
eccentricity on their own timescales, collision is enhanced and
eccentricity  will be damped to forced values. Thus,
oscillation of eccentricity damps quickly in wide models. 

We performed a simulation of narrow disk models to see how this behavior is
affected by the mass of planetesimals. The bottom panel of Figure \ref
{time_e_e25_alpha_multi} shows the time evolution of eccentricity of
model 8. The behavior of model 8 is almost the same as that of model
5. Thus, the number of planetesimals or mass of each planetesimal do
not change the result. In  narrow disk models, the convergence to
the forced eccentricity vector never occur.  

\subsubsection{Time evolution of proper eccentricity}
\begin{figure}

\plotone{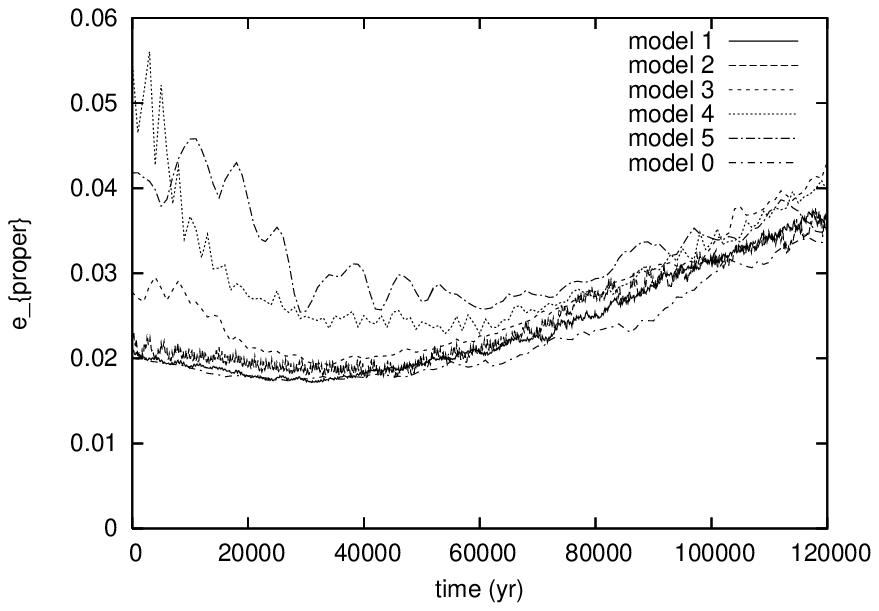}

\caption{
The time evolution of average of  proper eccentricity. the proper
eccentricity is averaged for planetesimals with $0.95 ~\mathrm{AU} < a
< 1.05 ~\mathrm{AU}$.
Solid, long-dashed, short-dashed, dotted, and upper dot-dashed curves show the results of
models 1 - 5, respectively. Lower dot-dashed curve shows the result of
model 0.
}
\label{e_eforced_global}

\end{figure}

Time evolution of proper eccentricity, ({\it i.e.,} distance from the
forced eccentricity in k-h plane) is shown in figure \ref 
{e_eforced_global}. The proper eccentricity is averaged for
planetesimal with semi-major axis between $0.95~\mathrm{AU}<a<
1.05~\mathrm{AU} $. In  circular models (model 3, 4, and 5), the
proper eccentricity quickly decreases through collisional damping and
gravitational relaxation. This means that the eccentricity vector of
the planetesimals converge to the forced eccentricity vector. In
forced models (model 1, 2) and single star model (model 0), on the
other hand, the proper eccentricity does not change significantly in early
stage of the simulation (before 40,000 years). 	
The convergence	in model 5 ($\alpha$ Cen  model) is slower than that
in the other two circular models and before convergence the planetary
accretion has almost finished. This is because of the small angular
velocity {\it A} of $\alpha$ Cen system. However, if we adopt real
radii, the convergence would probably occur before the accretion
completes. The time evolution of proper eccentricity after the
convergence is similar to that in the forced models and the single
star model. The proper eccentricity increases in the late stage due to 
the viscous stirring.

\begin{figure}

\plotone{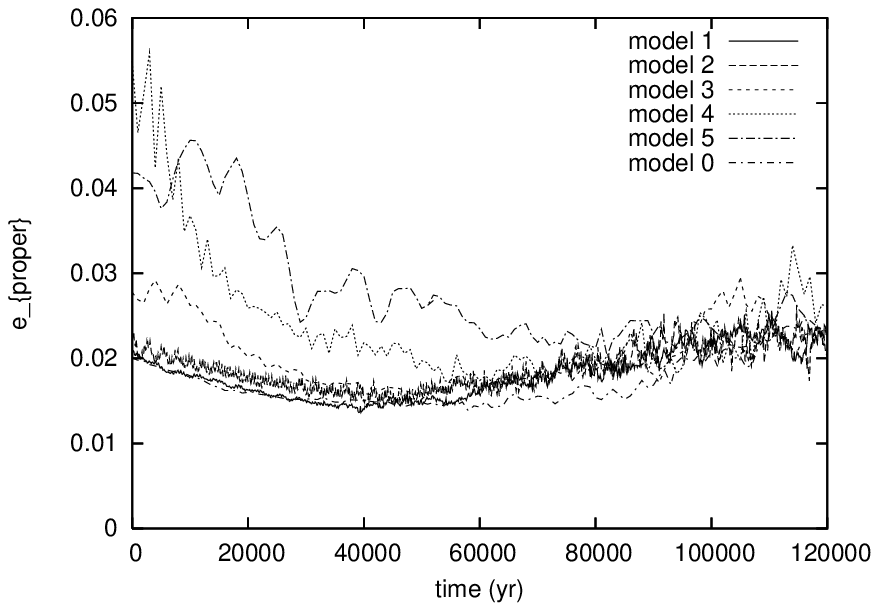}

\caption{
The same as figure \ref {e_eforced_global} but mass-weighted averages
are shown.
}

\label{e_eforced_mass}

\end{figure}

Figure \ref {e_eforced_mass}  shows time evolution of the mass
weighted average of proper eccentricity for the same radial range as
in figure  \ref {e_eforced_global}. In the early stage of the
simulation, its behavior is similar to that of the simple average in
figure \ref {e_eforced_global}. However, in the late stage, the
increase is slow because of dynamical friction to seeds of protoplanet.
In single star systems, the eccentricity of seeds of protoplanet
becomes small ({\it i.e.}, the eccentricity vectors converge to the origin of
 k-h plane).  On the other hand, in binary systems,  the
eccentricity vectors converge to the forced eccentricity and the
evolution of proper eccentricity is similar to the evolution of
eccentricity in the single star system. This means that the orbit of
planetesimals does not become circular as in single star system
but become eccentric in binary systems and their pericenter aligns to
the pericenter of companion star.
  
The distance between planetesimals on k-h plane determines the
collision velocity. Thus, the convergence reduces the collision
velocity between planetesimals. In model alpha, for example, the
maximum distance between planetesimals could becomes 0.08 by secular
oscillation if the mutual gravitational effect is neglected. It
corresponds to the collision velocity of 3000-4000  $\mathrm{m~s^{-1}}$
at 1 AU.  Collision with this velocity is destructive even for the
mass of the planetesimal of 
about $1.4 \times 10^{24}$ g (its escape velocity is about 600
$\mathrm{m~s^{-1}}$). On the other hand, The collision velocity is
reduced to about 500-1000 $\mathrm{m~s^{-1}}$ with this convergence.
As we mentioned above, gas drag
affects the eccentricity dispersion of the planetesimals ({\it i.e.,} 
gas drag can not be negligible in this sense). The simulations in
which gas drag and mutual gravitational effects are taken
into account is required to determine the precise value of the
collision velocity. 

\section{Conclusion and Discussion}
We have performed the simulations of formation of protoplanets from
massive planetesimals (with  radii of 100km-500km) in close binary
systems, for which gas drag effect is negligible and the
coupling of the secular perturbation and gravitational interaction
is important. 

We found that eccentricity vectors of planetesimals quickly
converge to the forced eccentricity if the initial disk model is
sufficiently wide in radius and secular perturbation is
sufficiently strong. The eccentricity vectors of planetesimals which
have heavier mass than the other planetesimals  converge
to the forced eccentricity more strongly. This convergence results in
orbital phasing and it reduces value of collision velocity to
less than the value predicted in studies with  restricted three-body
approach.  
Runaway growth and oligarchic growth also occur in binary system much
in the same way as isolated star system at least in the late stage
of the formation. 
The final configuration of protoplanets is not different between
close binary systems and isolated star systems. 

Our simulations, however, have following limitations. (1) We
underestimated effect of secular perturbation. The {\it f}-fold
change of radius increases frequency of collision and accelerates
planetary accretion. This acceleration of the formation process causes
relative under-estimate of secular perturbation. We plan to
perform the simulation with real radii of planetesimals to see if this
effect would make any difference. (2) We investigated only three binary
systems. More simulations of various binary systems are
required to determine quantitative relationship between the
convergence of eccentricity vectors  and strength of secular
perturbation. (3) We neglected effect of gas drag. This is okay for
the late stage which we studied. In earlier stage, however, gas drag
is clearly important. We plan to calculate the equilibrium state of
gas disk under the secular perturbations and perform the
simulations which include the gas effect in future works.

\section *{Acknowledgments}
We express our sincere gratitude to Masaki Iwasawa and Keigo Nitadori
for useful comments on our {\it N}-body code.  We also thank Eiichiro
Kokubo and Satoshi Inaba for many useful suggestions. We thank the
anonymous referee whose critical comments helped us greatly to make
the paper clear. This research is partially supported by the Special
Coordination Fund for Promoting Science and Technology (GRAPE-DR
project), Ministry of Education, Culture, Sports, Science and
Technology, Japan.

\end{document}